\documentclass[aps,preprint,superscriptaddress]{revtex4}

\usepackage{graphicx}
\usepackage{amsmath}
\usepackage{amssymb}

\newcommand{\br}{\mathbf{r}}
\newcommand{\bp}{\mathbf{p}}
\newcommand{\atomes}[1]{\mathrm{#1}}
\newcommand{\unites}[1]{\; \mathrm{#1}}   

\begin{document}
\title{Single and double ionization of magnesium by electron impact:  A classical study }
%\date{\today}

\author{J. Dubois}
\affiliation{Aix Marseille Univ, CNRS, Centrale Marseille, I2M, Marseille, France}
\author{S. A. Berman}
\affiliation{Aix Marseille Univ, CNRS, Centrale Marseille, I2M, Marseille, France}
\affiliation{School of Physics, Georgia Institute of Technology, Atlanta, Georgia 30332-0430, USA}
\author{C. Chandre}
\affiliation{Aix Marseille Univ, CNRS, Centrale Marseille, I2M, Marseille, France}
\author{T. Uzer}
\affiliation{School of Physics, Georgia Institute of Technology, Atlanta, Georgia 30332-0430, USA}

\begin{abstract}
We consider electron impact-driven single and double ionization of magnesium in the 10-100 $ \unites{eV}$ energy range. Our classical Hamiltonian model of these $(e,2 e)$ and $(e,3 e)$ processes sheds light on their total cross sections and reveals the underlying ionization mechanisms. Two pathways are at play in single ionization: Delayed and direct. In contrast, only the direct  process is observed in double ionization, ruling out the excitation-autoionization channel. We also provide evidence that the so-called TS2 (Two-Step~2) mechanism predominates over the TS1 (Two-Step~1) mechanism, in agreement with experiments. 
\end{abstract}

% PhySH
% Atomic and Molecular Collisions
% Hamiltonian Systems
% Chaos and Nonlinear Dynamics

\maketitle
%\tableofcontents

\section{Introduction}
\begin{figure}
\centering
\includegraphics[width=0.8\textwidth]{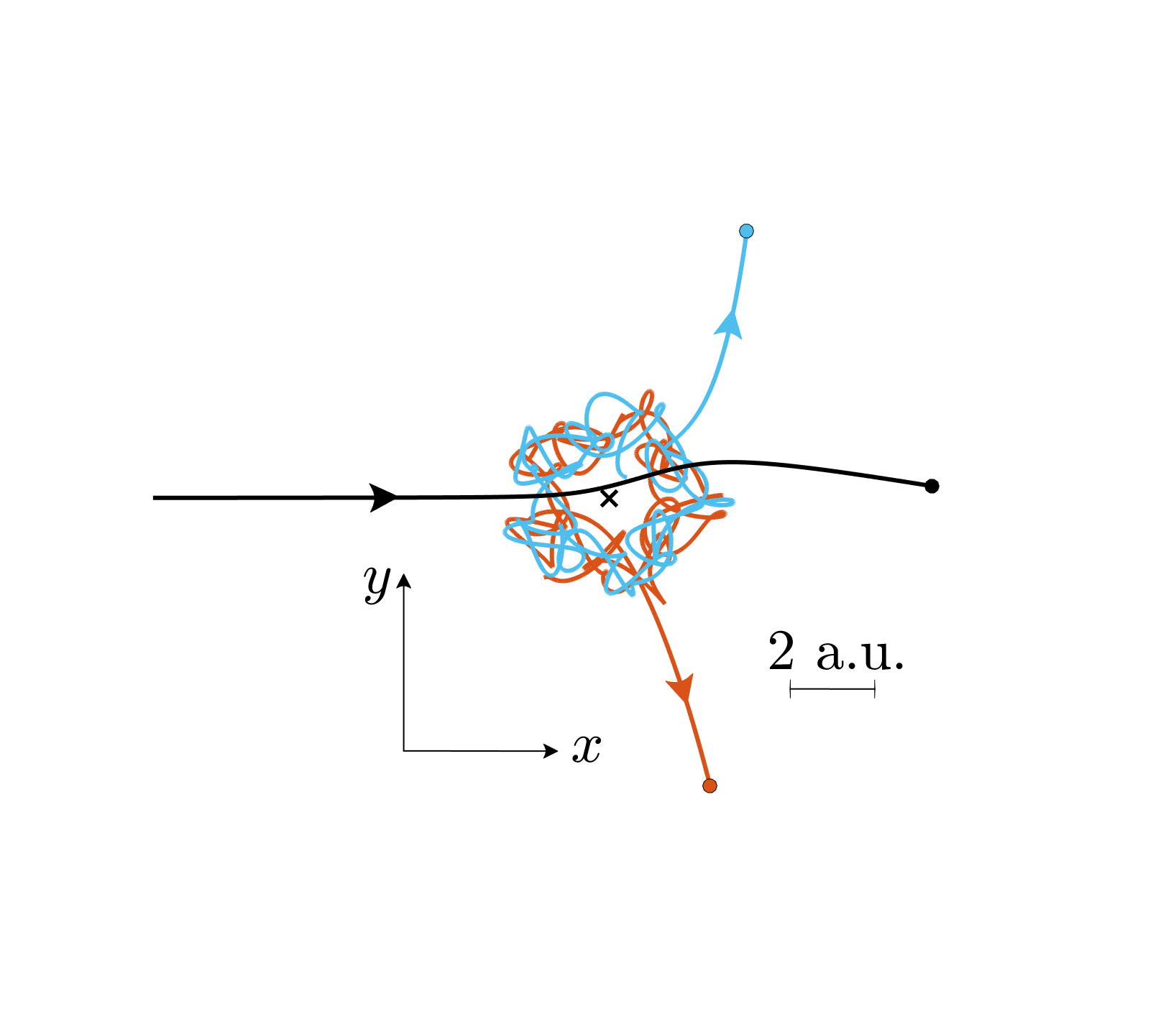}
\caption{(Color online) Trajectory of each of the three electrons in configuration space during a double ionization of a target by electron impact. The impact electron is in black, the two electrons of the target are in gray (red) and light gray (blue). The position of the ionic core is identified by a cross.}
\label{fig:trajectoire}
\end{figure}
A complete understanding of the electronic dynamics and the structure of atoms and molecules has been the focus of many theoretical and experimental studies. A common technique to extract information from such small entities is to perturb the system. The perturbation must be strong enough to compete with the strong Coulomb interactions inside atoms. Typical methods are the application of intense laser pulses in strong field physics or the impact of particles in atomic collision physics. The products of such processes are emitted light or multiple ionization, which are then measured in experiments. Here, we consider multiple ionization by electron impact to recover some information on the energy exchanges between electrons, the electronic dynamics inside atoms, and the structure of the target. Electron impact processes have a wide range of applicability, ranging from modeling in fusion plasmas (for a review, see Ref.~\citep{Bolt2002}) to astrophysics (as in planetary upper atmospheres)~\cite{Kallman2007}.  
\par
In electron impact experiments [so-called $(e, n e)$ processes], an electron beam is directed on a target gas of atoms or ions. The gas beam and the electron beam move in perpendicular directions (crossed-electron-beam-- fast-atom-beam method). The target is usually in the ground state. If the impact energy is in a certain energy range, some of the target atoms ionize. The scattered and ejected electrons are detected in coincidence so as to properly compute the various total and differential cross-sections~\citep{Ehrhardt1986}. The $(n-1)$-tuple ionization of a target $X$ by electron impact is described by
\begin{equation}
X + e^- \to X^{(n-1)+} + n e^- ,
\end{equation}
where $X^{(n-1)+}$ is the ion with charge $(n-1)e$. In this article we focus on single and double ionization, i.e. $n=2,3$. The basic mechanisms at play in the ionization of atoms are classified in two main categories: 
Direct ionization processes, where the ionization occurs immediately after the impact, and indirect ionization processes, such as excitation-autoionization, where the ionization occurs some time after the impact. The following energy regions associated with different values of the impact energy are sequentially defined~\citep{Ehrhardt1986}:
\begin{itemize}
\item[$\bullet$]  The low energy region, where only the outer shell is involved in the ionization processes. 
\item[$\bullet$] The intermediate energy region, where the inner shell can also be involved in the ionization processes.
\item[$\bullet$] The high energy region, where very few atoms are ionized because interaction times are too brief. 
\end{itemize}
\par
Between the low energy region and the intermediate energy region for $\atomes{Mg}$, Okudaira \textit{et al} \citep{Okudaira1970} (1970), McCallion \textit{et al} \citep{McCallion1992a} (1992), and Boivin \textit{et al} \citep{Boivin1998} (1998), have reported a ``discontinuity'' in the total cross section of the double ionization, i.e. $(e,3e)$ process, in the $40-60 \unites{eV}$ energy range, evidencing different ionization channels at play. This so-called discontinuity is a crossover between the direct and the indirect processes~\citep{Jonauskas2014}. The double ionization crossover has been observed for a wide variety of targets other than Mg: Be-like ions like ${\rm B}^+$~\citep{Shevelko2005,Pindzola2011}, Ba~\citep{Dettmann1982,Jha1994}, Ca and Sr~\citep{Okudaira1970,Chatterjee1984}, ${\rm Ar}$ and ${\rm Xe}$ ions~\citep{Pindzola1984,McCallion1992b} (see also Ref.~\citep{Freund1990} for other targets). 
\par
Reference~\citep{Peach1970} attributes this discontinuity to the rise of the Auger effect in the intermediate energy region. For $\atomes{Mg}$, the Auger effect is predicted to start at $55.8 \unites{eV}$, and for impact energy higher than $60 \unites{eV}$, this effect dominates the direct ionization processes. The indirect processes are usually very well described theoretically with good quantitative agreement with experimental data. Perturbative methods provide good results for indirect processes involving tightly bound inner shell electrons of heavy atoms. In contrast, the direct double ionization is more intricate to handle theoretically since it is driven by strong correlations between the three electrons (see Refs.~\citep{Pindzola2004,Pindzola2009}).  
\par
The double ionization problem is too complex to be treated fully quantum mechanically, and cannot be treated by sequential approximations or perturbative methods. Reduced non-perturbative methods have been designed to reproduce quantitatively experimentally observed cross sections. For example, time-dependent close coupling (TDCC)~\citep{Pindzola2007}, R-matrix with pseudostates (RMPSs)~\citep{Ballance2007}, and convergent close coupling (CCC)~\citep{Bray2002}, are tested and benchmarked (see, e.g., Refs.~\citep{Jiao2008,Jha2002,Pindzola2009}). 
Cross sections for Mg have been measured experimentally since the 70s and analyzed theoretically ever since (see Refs.~\citep{Okudaira1970,Vainshtein1972,Karstensen1978,Chatterjee1982,McCallion1992a,Boivin1998,Jha2002,Pindzola2009}). 
\par
When it comes to gaining qualitative understanding and uncovering mechanisms, classical trajectory methods have an excellent track record \citep{Becker2008}. Uncovering the influence of the strong electron-electron interaction on the ionization processes by classical-mechanical means is the main focus of the present work. We consider a target with two strongly coupled electrons, such as Mg, and consequently we consider the fully coupled four body Coulomb problem. Figure~\ref{fig:trajectoire} shows a typical double ionization trajectory in configuration space, computed from the Hamiltonian model proposed in Sec.~\ref{sec:model}. An impact electron is sent from the left, far away from the target, in the direction of the latter, with a given impact energy $\epsilon_0$. As a result of the three electron interaction, the impact electron is scattered and the two target electrons are ejected from the target. The specific objective of this article is to understand classically the mechanisms behind the total cross section of the single and the double ionization of $\atomes{Mg}$. Our focus is on the direct double ionization where only the two 3s electrons are involved. To this end, we propose a two-active electron model with a soft-Coulomb interaction potential. The strong electron-electron interaction inside the target atom is fully taken into account in our model. The main advantage of a classical model is two-fold: First, it is easy to integrate numerically, and second, classical trajectories allow one to visualize the electronic dynamics in phase space and understand the ionization mechanisms and their occurrence as parameters (such as impact energy or target) are varied. The proposed classical model is complementary to quantum approaches. Despite the simplicity of our model, we also provide evidence that the so-called Two-Step~2 mechanism (TS2, in which the impact electron hits both 3s outer shell electrons), predominates over the Two-Step~1 mechanism (TS1, in which the impact electron hits only one 3s electron), in agreement with experiments. 
\par
The article is organized as follows: In Sec.~\ref{sec:model}, we introduce the classical two-active electron model, and we compute and discuss the single and double ionization probability curves as functions of the impact energy $\epsilon_0$. The mechanisms behind these curves are identified and analyzed in Sec.~\ref{sec:mechanisms}. The probability of each mechanism as a function of the impact energy $\epsilon_0$ is computed and compared with the literature.  

\section{Our model and its ionization probability curves \label{sec:model}}
In this section we present the classical Hamiltonian model we choose for the description of the $(e,2 e)$ and $(e,3 e)$ processes. We consider a two-active electron model for $\atomes{Mg}$, describing the dynamics of the two most loosely bound electrons, the 3s electrons. Using this model, we compute and discuss the single and double ionization probabilities as a function of the impact energy $\epsilon_0$.

\subsection{The model}
We consider a $d$-dimensional configuration space, ${\mathbb R}^d$ with $d=1,2$ or $3$. The positions and the canonical momenta of the two active electrons of the target are denoted $\br_k$ and $\bp_k$ respectively, with $k=1,2$. We consider a static ionic core (Born-Oppenheimer approximation) located at the origin of the configuration space. We have checked that all the results we present below are the same with and without the Born-Oppenheimer approximation, in the energy range we consider. In atomic units, the Hamiltonian of the isolated target reads
\begin{eqnarray}
H_T &=& \frac{|\bp_1|^2}{2} + \frac{|\bp_2|^2}{2} - \frac{2}{\sqrt{|\br_1|^2 + a^2}}
- \frac{2}{\sqrt{|\br_2|^2 + a^2}}\nonumber\\ && \qquad + \frac{1}{\sqrt{|\br_1 - \br_2|^2 + b^2}} .
\label{eq:Hamiltonian_atom}
\end{eqnarray}
Initially, the target is in its ground state of energy $\mathcal{E}_g$ defined as the sum of the first two ionization potentials, i.e. $\mathcal{E}_g =\mathcal{E}_1+\mathcal{E}_2= -0.83 \unites{a.u.}$ for $\atomes{Mg}$. The charged particle interaction we use is the soft-Coulomb potential~\citep{Javanainen1988} which is widely used in strong field atomic physics~\citep{Becker2008}. The softening parameters $a$ and $b$, which control the electron-ion and the electron-electron interaction respectively, are chosen such that the ground state energy surface is not empty and there is no self-ionization. For $\atomes{Mg}$, these conditions are satisfied for $a = 3 \unites{a.u.}$ and any $b$, so unless otherwise specified, $b = 1 \unites{a.u.}$ The value of $b$ does not have a qualitative influence on the ionization mechanisms, as we show in Sec.~\ref{sec:robustness}.
\par
The position and the canonical momentum of the impact electron are denoted $\br_0$ and $\bp_0$. The dynamics of the impact electron with its interaction with the target is described by the following Hamiltonian
\begin{eqnarray*}
H_I &=& \frac{|\bp_0|^2}{2} - \frac{2}{\sqrt{|\br_0|^2 + a^2}}\nonumber \\
&&  +  \frac{1}{\sqrt{|\br_0 - \br_1 |^2 + b^2}} + \frac{1}{\sqrt{|\br_0 - \br_2|^2 + b^2}},
\end{eqnarray*}
such that the total Hamiltonian is
\begin{equation}
H  = H_T + H_I  .
\label{eq:Hamiltonian}
\end{equation}
Initially, the impact electron has a given kinetic energy, denoted by $\epsilon_0$, and its position is far away from the target. 
This Hamiltonian system has $3 d$ degrees of freedom. We notice that Hamiltonian~(\ref{eq:Hamiltonian}) is invariant under time translations and rotations (for $d \geq 2$). Consequently, the total energy and the total angular momentum of the system are conserved, corresponding to $d$ conserved quantities. For any $d$, the dynamics could potentially exhibit chaotic behavior. Because of energy conservation, at any time, Hamiltonian~(\ref{eq:Hamiltonian}) satisfies
\begin{equation}
H = \epsilon_0 + \mathcal{E}_g .
\label{eq:total_energy}
\end{equation} 

\subsection{The probability curves}

In order to compute single and double ionizations, we count the number electrons at a distance greater than $L$ from the target at the end of the simulation. In practice we choose $L = 150 \unites{a.u.}$ in order to ensure that the interaction between the ionic core and an electron at a distance $L$ from the target is negligible. The integration time is $t_f = 800 \unites{a.u.}$, i.e. a total integration time of $t_f - t_i$.
\par
We initiate the three electron dynamics at a time $t_i = - 100 \unites{a.u.}$, launching the impact electron in the direction of the target in the positive $x$-direction (see Fig.~\ref{fig:trajectoire}) with a kinetic energy $\epsilon_0$. Consequently, the initial condition of the impact electron is   
$$
\bp_0 (t_i)  = \sqrt{2 \, \epsilon_0} \: \mathbf{\hat{x}},
$$
and
$$
\br_0 (t_i) = t_i \: \bp_0 (t_i),
$$
such that $t=0$ corresponds to the moment when the impact electron reaches the origin of the configuration space in the absence of target. If $d \geq 2$, this configuration corresponds to zero impact parameter. We notice that considering a range of impact parameters has no qualitative influence on the ionization probability curves and on the various processes at play. Its main quantitative effect is to decrease the ionization probabilities for increasing impact parameters. As we show in Sec.~\ref{sec:robustness} we choose $y_0 (t_i) = 0$ in order to maximize the ionization probability. The two-active electron target is initiated with a microcanonical distribution with energy ${\cal E}_g$ as in Ref.~\citep{Mauger2012}. 
\begin{figure}
\centering
\includegraphics[width=0.8\textwidth]{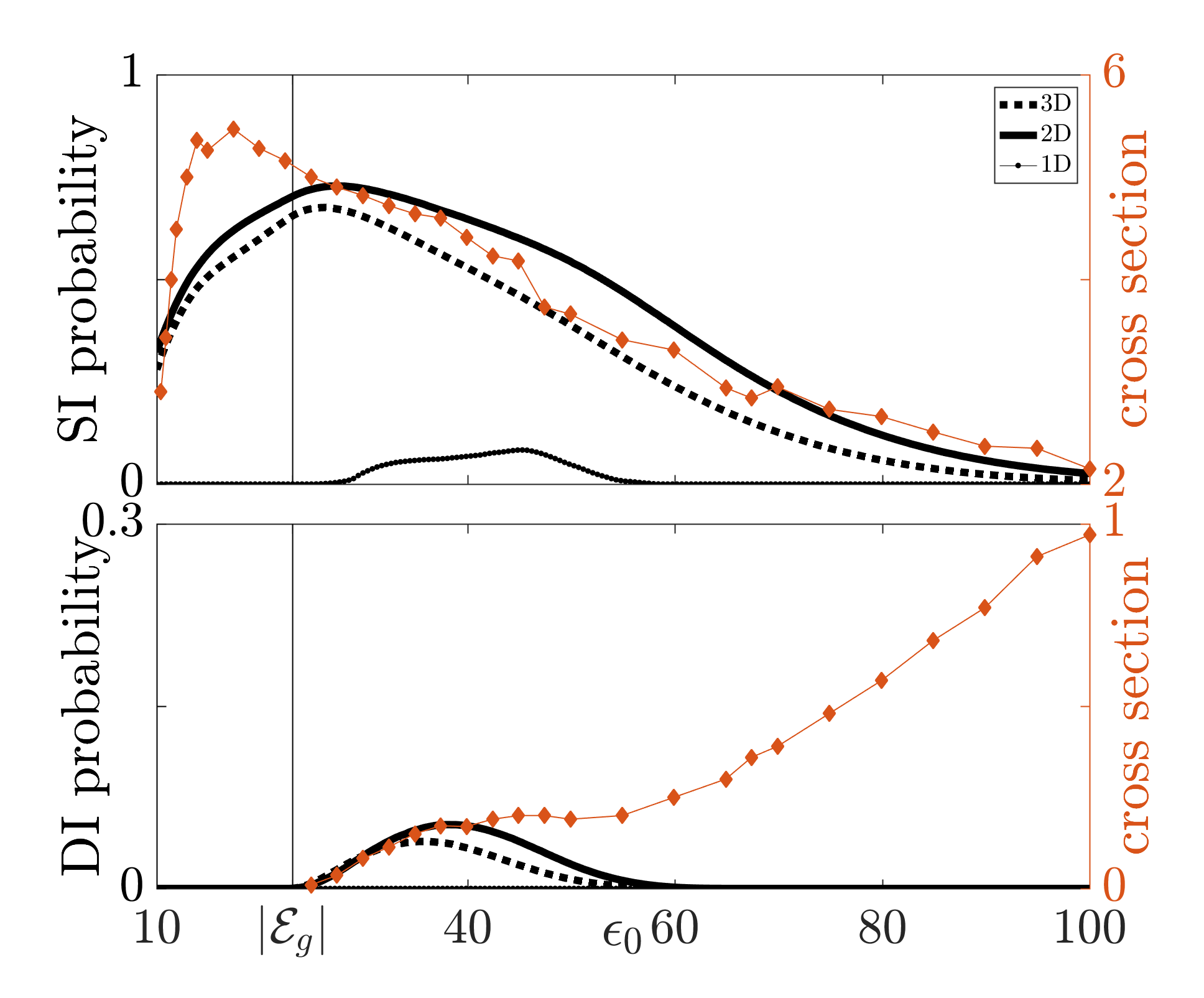}
\caption{(Color online) In the upper and the lower panel, respectively the single ionization (SI) and the double ionization (DI) probability of Hamiltonian~(\ref{eq:Hamiltonian}), for 1D, 2D and 3D models (respectively $d=1,2,3$), measured at $t_f = 800 \unites{a.u.}$ The vertical vertical line is at $\epsilon_0 = | \mathcal{E}_g | $. The solid diamonds are the experimental results of Ref.~\citep{McCallion1992a}. Cross sections in $\unites{Mb}$ ($1 \unites{Mb} = 10^{-18} \unites{cm}^2$) and $\epsilon_0$ in $\unites{eV}$.}
\label{fig:SI_DI}
\end{figure}
Figure \ref{fig:SI_DI} shows, respectively in the upper and the lower panel, the single and the double ionization probabilities of Hamiltonian~(\ref{eq:Hamiltonian}), for $d = 1,2,3$, as a function of the impact energy $\epsilon_0$. These curves are obtained by generating $10^7$ trajectories for each value of $\epsilon_0$.
\par
All single ionization curves display the same qualitative behavior: An increase of single ionization with increasing impact energy, followed by a decrease for larger values of the impact energy. For low values of $\epsilon_0$, there is not enough energy to be transferred to the target, and for higher values of $\epsilon_0$, the impact electron is too fast to transfer energy to the target. This leaves a rather narrow interval for single ionization, here, $\epsilon_0 \in [10 , 100] \unites{eV}$, in qualitative agreement with the experimental data of Ref.~\citep{McCallion1992a}. In the upper panel of Fig.~\ref{fig:SI_DI}, we observe that for $d = 1$, the single ionization probability is significantly smaller than the ones for $d=2$ and $d = 3$, and is non-zero only in a too narrow range $\epsilon_0 \in [ | \mathcal{E}_g | , 60] \unites{eV}$, in contrast with the ionization probability obtained for $d=2$ and $d=3$. We observe that the maximum of single ionization for $d=2$ and $d=3$ is obtained for $\epsilon_0$ close to $ | \mathcal{E}_g |$ which is qualitatively consistent with the experimental data of Ref.~\citep{McCallion1992a}. In Sec.~\ref{sec:mechanisms}, we will show that the location of this maximum depends on the chosen integration time of the simulation.
\par
In the lower panel of Fig.~\ref{fig:SI_DI}, we observe that there is no double ionization for any value of the impact energy in $d = 1$, in stark disagreement with $d>1$ and with the experimental data. For both single or double ionization no significant differences are observed between the two cases $d=2$ and $d=3$. The double ionization probability curves for $d=2$ and $d=3$ are non-zero in the range $\epsilon_0 \in [|\mathcal{E}_g| , 60] \unites{eV}$, and reach a maximum for $\epsilon_0 = 40 \unites{eV}$. These curves resemble the total cross section of Ref.~\citep{McCallion1992a} in the low energy region. Even though the $d=1$ model is more easily analyzed due to the low dimensionality of its phase space, it needs to be discarded since it leads to erroneous conclusions. Indeed, given that the two 3s electrons are aligned, they are ejected in the same direction. Then, the electron-electron repulsion pushes at least one electron (the closest to the core) back to the core. For $d \geq 2$, the electron-electron repulsion moves the electrons in opposite direction in the transverse plane, opening the double ionization channel. For practical purposes, we consider the case $d=2$ for the analysis of the impact ionization dynamics. 
\par
The striking feature of the experimental double ionization curve is the presence of a bump in the low energy part. This has been referred to as a ``discontinuity'' in Refs.~\citep{McCallion1992a,Boivin1998,Jha2002}, but here we prefer to call it a ``knee'' in analogy to the knee in the double ionization of atoms by strong laser pulses~\citep{Becker2008}. The double ionization curve obtained for $d=2$ and the experimental one are significantly different for large values of the impact energy. If $\epsilon_0 > 50 \unites{eV}$, the direct ionization processes' contribution becomes smaller with increasing impact energy, and the inner shell contribution cannot be neglected. Our two-active electron model captures the first part of the knee, that corresponding to the outer shell contribution, which is due to some three-electron processes which we analyze in what follows. The second part of the knee corresponds to the inner shell contribution, which is not taken into account in our inherently outer shell model. In order to reproduce the entire experimental total double ionization cross section, one should add a third electron (with an energy given by the third ionization potential) in the model. This is beyond the scope of the present work. 

\section{Mechanisms of single and double ionization \label{sec:mechanisms}}
In this section we investigate the mechanisms involved in the single and double ionization of the two-active electron model~(\ref{eq:Hamiltonian}) for $d=2$. We study where and how often these mechanisms occur in phase space. We show the contribution of each mechanism building up the single and double ionization probability curves. 

\subsection{Single ionization mechanisms}
\begin{figure}
\centering
\includegraphics[width=0.8\textwidth]{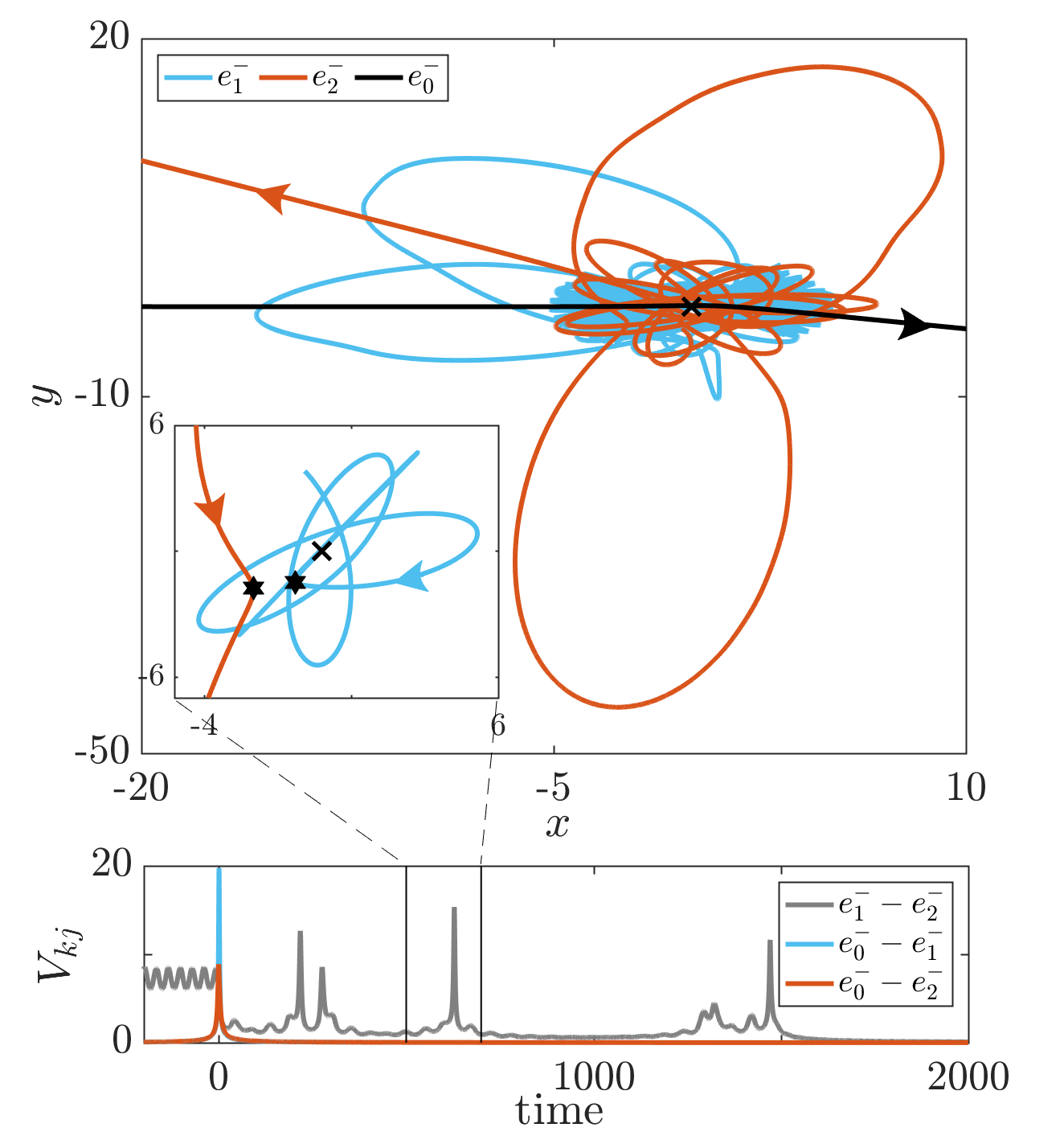}
\caption{(Color online) Delayed single ionization. Upper panel: Trajectory of each of the three electrons in configuration space $(x,y)$. Only the trajectory for positive times (i.e. after the impact) is represented. The black trajectory corresponds to the impact electron. The gray (red) and the light gray (blue) trajectories correspond to the two electrons of the target. In the inset the trajectory of the electrons of the atom for $t \in [500,700] \unites{a.u.}$ is represented. The instant of the collision is indicated by a pair of stars. The position of the ionic core is indicated by a cross. Lower panel: Corresponding $V_{kj}$, defined by Eq.~(\ref{eq:interaction}), as a function of time for $(k,j) = (0,1)$, $(0,2)$, and $(1,2)$. The impact energy is $\epsilon_0=40 \unites{eV}$. Here $V_{kj}$ in $\unites{eV}$ and all other quantities in atomic units.}
\label{fig:delayed}
\end{figure} 
\begin{figure}
\centering
\includegraphics[width=0.8\textwidth]{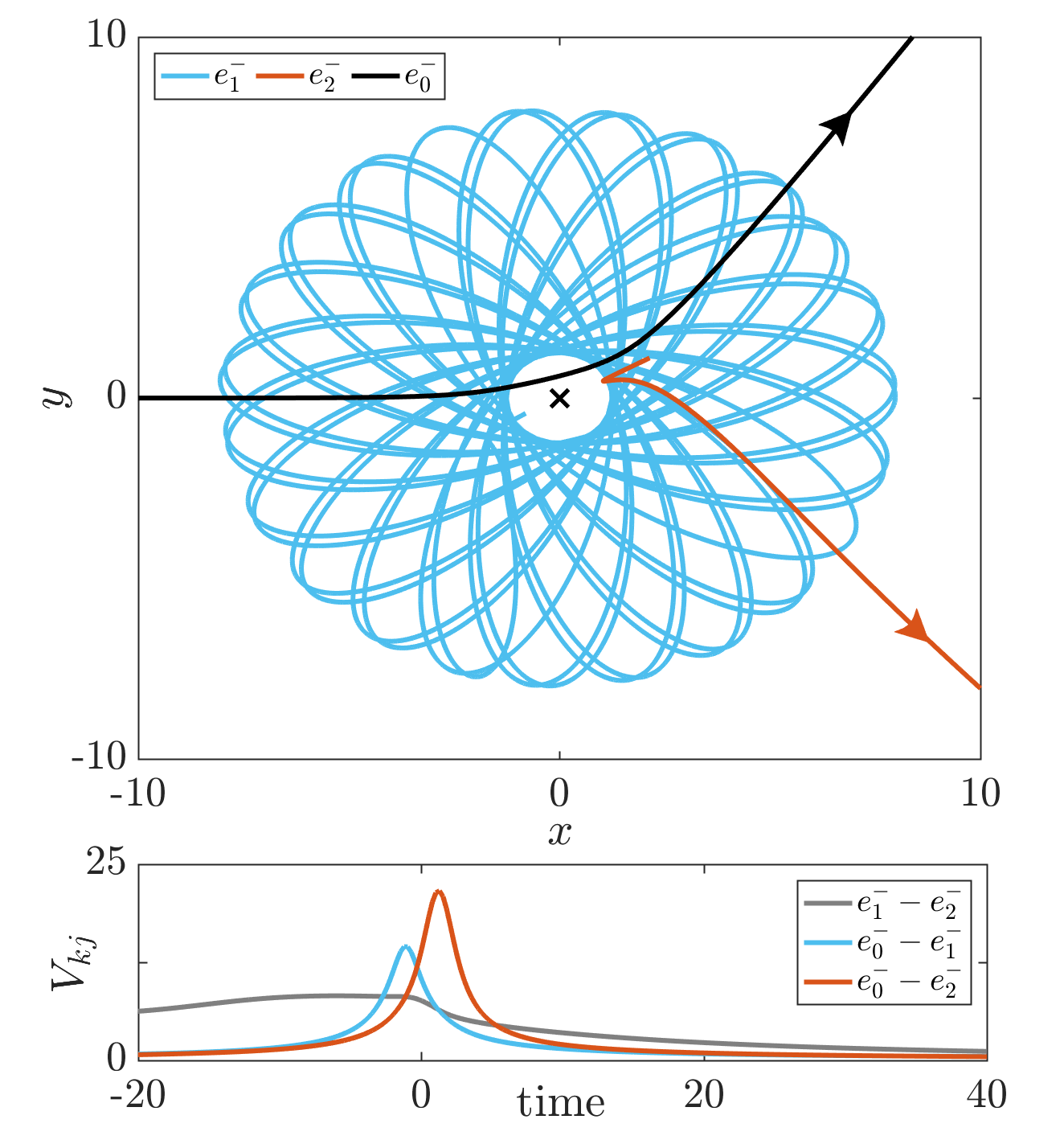}
\caption{(Color online) Direct single ionization. Upper panel: Trajectory of each of the three electrons in configuration space $(x,y)$. Only the part of the trajectory after the impact is represented. The black trajectory corresponds to the impact electron $e_0^-$. The gray (red) and the light gray (blue) trajectories correspond to the two electrons of the target $e_1^-$ and $e_2^-$. The position of the ionic core is indicated by a cross. Lower panel: Corresponding $V_{kj}$, defined by Eq.~(\ref{eq:interaction}), as a function of time for $(k,j) = (0,1)$, $(0,2)$, and $(1,2)$. The impact energy is $\epsilon_0=40 \unites{eV}$. Here $V_{kj}$ in $\unites{eV}$ and all other quantities in atomic units.}
\label{fig:direct}
\end{figure} 
Two distinct single ionization mechanisms have been identified  in this process: Direct single ionization and delayed single ionization. In the upper panels of Figs.~\ref{fig:delayed} and \ref{fig:direct} we represent a typical three-electron trajectory in configuration space for each kind of single ionization for $\epsilon_0=40\unites{eV}$. In the lower panels, the interaction potentials 
\begin{equation}
V_{kj} (t) = \frac{1}{\sqrt{|\br_k (t) - \br_j (t)|^2 + b^2}} ,
\label{eq:interaction}
\end{equation}
between the electrons $k$ and $j$ for $(k,j) = (0,1)$, $(0,2)$ and $(1,2)$ are represented as functions of time. The peaks indicate collisions between two electrons. The maximum of the interaction energy~(\ref{eq:interaction}) is $1/b = 27.2 \unites{eV}$, and occurs when the two electrons overlap.
\par
Figure~\ref{fig:delayed} represents a delayed single ionization. We observe some collisions between the impact electron and the target electrons around $t=0$, as expected. Then the impact electron leaves the target region after losing some energy to excite the target. The two target electrons stay bounded to the ionic core up to $t=1500  \unites{a.u.}$ and collide with each other several times, exchanging energy. We notice that most of the time the two target electrons are far away from each other after the impact, interacting only at some specific times by collisions. Finally, one of the target electrons ($e_1^-$ in Fig.~\ref{fig:delayed}) leaves the target region while the other one remains bounded. 
%During such a process, the red (gray) electron can be ejected, with the same probability, in any direction.
\par
Figure~\ref{fig:direct} represents a direct single ionization. We observe one peak on each of the impact-target electron interaction curves near $t=0$ where the impact occurs. The first interaction is between the impact electron and $e_1^-$, which does not leave the ionic core but is excited. Then, the impact electron collides with $e_2^-$. The interaction is stronger than the first one, and makes $e_2^-$ ionize. 
\par
In the case of a delayed single ionization, the time it takes the atom to ionize after excitation strongly depends on the initial conditions, and can be arbitrarily long. In order to gain insight into the dynamics, we would like to address the following questions: What are the conditions leading to one or the other single ionization mechanism? Is any mechanism  favored by the dynamics? These questions are addressed by examining the organization of the dynamics in phase space.   
\begin{figure}
\centering
\includegraphics[width=0.8\textwidth]{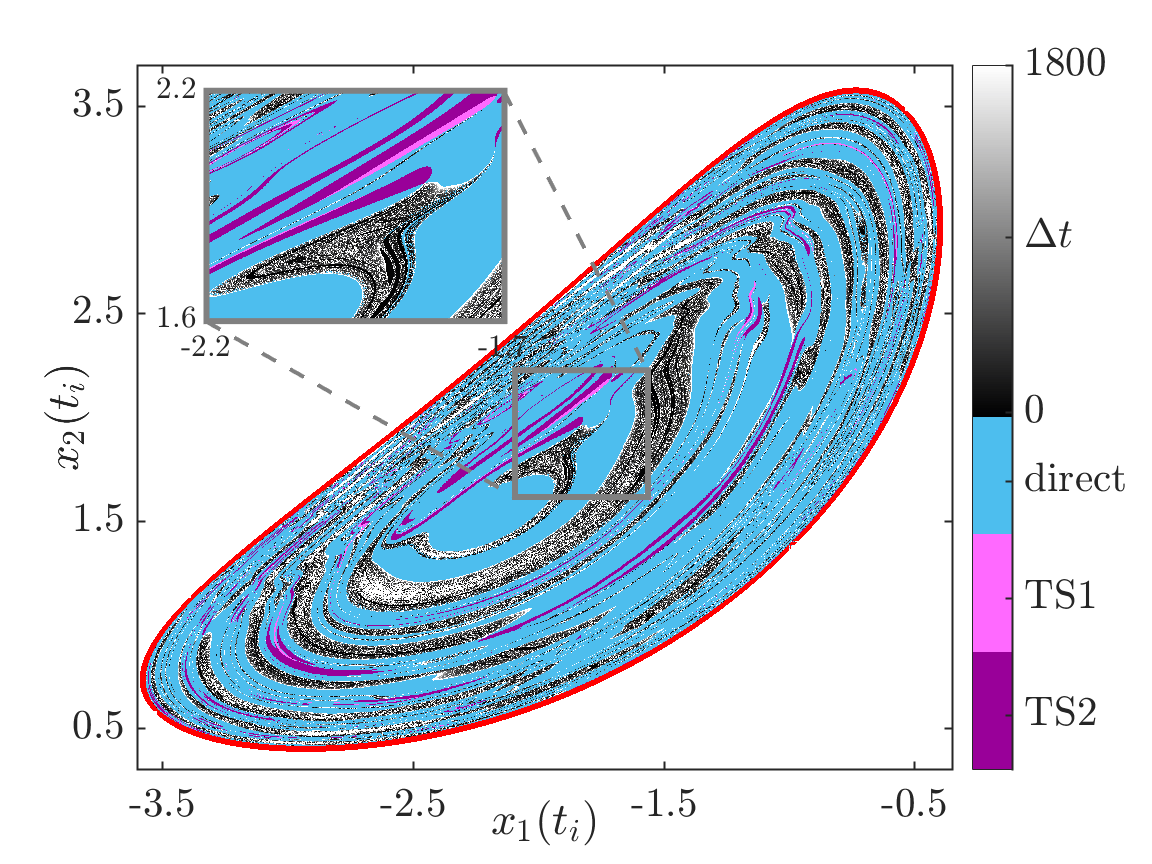}
\caption{
%
%(Color online) Slice of the phase space of the isolated target at time $t_i$ in the plane $(x_1(t_i),x_2(t_i))$ for $y_1(t_i) = y_2 (t_i) = 0$, $\bp_1(t_i) = (P/\sqrt{2}, -P/2)$ and $\bp_2(t_i) = (\sqrt{2} P/4,\sqrt{2} P / 4)$, with $P$ the solution of Eq.~(\ref{eq:def_P}). Outside the gray (red) contour line $P$ is not defined. Here $\epsilon_0 = 40 \unites{eV}$ and $t_i = - 500 \unites{a.u.}$ The mechanism involved for each initial condition after the impact is represented. For delayed single ionization, the time $\Delta t$ spent by the atom to lose an electron after the impact is represented. The white areas inside the contour zone represent the initial conditions where the atom does not ionize. All quantities in atomic units.
%
(Color online) Ionization scenario as a function of the initial conditions $(x_1(t_i),x_2(t_i))$ for $y_1(t_i) = y_2 (t_i) = 0$, $\bp_1(t_i) = (P/\sqrt{2}, -P/2)$ and $\bp_2(t_i) = (\sqrt{2} P/4,\sqrt{2} P / 4)$, with $P$ the solution of Eq.~(\ref{eq:def_P}). Outside the gray (red) contour line $P$ is not defined. The impact energy is $\epsilon_0 = 40 \unites{eV}$, $t_i = - 500 \unites{a.u.}$, and $t_f = 2000\unites{a.u.}$ For delayed single ionization, the time $\Delta t$ spent by the atom to lose an electron after the impact is represented. The white areas inside the contour zone (red) represent the initial conditions where the atom does not ionize (or eventually ionize with a delay which is too long). All quantities in atomic units.
}
\label{fig:Manifold}
\end{figure}
\begin{figure}
\centering
\includegraphics[width=0.8\textwidth]{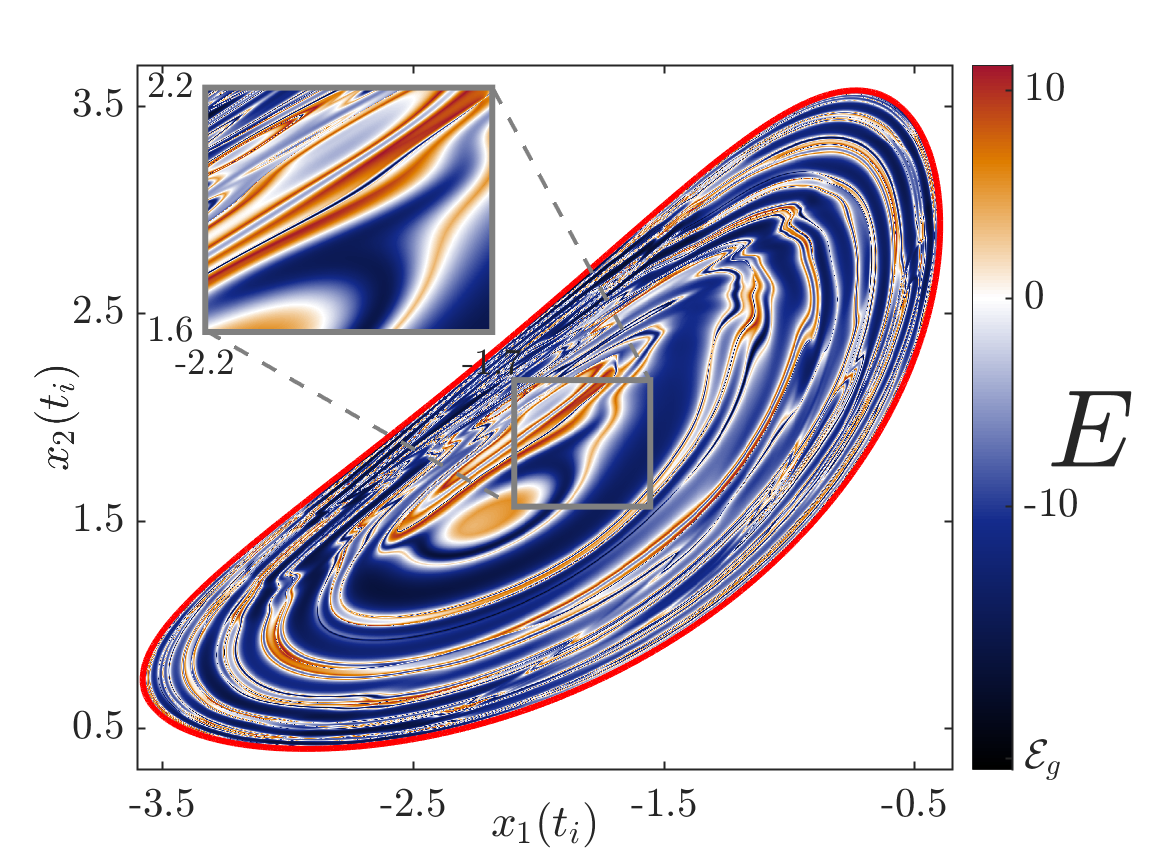}
\caption{
%
%(Color online) Slice of the phase space of the isolated target at time $t_i$ in the plane $(x_1(t_i),x_2(t_i))$ for $y_1(t_i) = y_2 (t_i) = 0$, $\bp_1(t_i) = (P/\sqrt{2}, -P/2)$ and $\bp_2(t_i) = (\sqrt{2} P/4,\sqrt{2} P / 4)$, with $P$ the solution of Eq.~(\ref{eq:def_P}). Outside the gray (red) contour line $P$ is not defined. Here $\epsilon_0 = 40 \unites{eV}$ and $t_i = - 500 \unites{a.u.}$ The energy of the atom $E$ after the impact time is represented. Here $E$ in $\unites{eV}$ and all other quantities in atomic units.
%
(Color online) Energy of the atom $E$ at $t_f = 2000\unites{a.u.}$ as a function of the initial conditions $(x_1(t_i),x_2(t_i))$ for $y_1(t_i) = y_2 (t_i) = 0$, $\bp_1(t_i) = (P/\sqrt{2}, -P/2)$ and $\bp_2(t_i) = (\sqrt{2} P/4,\sqrt{2} P / 4)$, with $P$ the solution of Eq.~(\ref{eq:def_P}). Outside the gray (red) contour line $P$ is not defined. The impact energy is $\epsilon_0 = 40 \unites{eV}$, and $t_i = - 500 \unites{a.u.}$ Here $E$ in $\unites{eV}$ and all other quantities in atomic units.
}
\label{fig:Energy}
\end{figure}
Figures \ref{fig:Manifold} and \ref{fig:Energy} depict a slice of the phase space for $\epsilon_0 = 40\unites{eV}$ in the plane $(x_1 (t_i) , x_2(t_i))$, for $y_1(t_i) = y_2(t_i) = 0$, $\bp_1(t_i) = (P/\sqrt{2}, -P/2)$ and $\bp_2(t_i) = (\sqrt{2} P/4,\sqrt{2} P / 4)$ where $P$ is such that 
\begin{eqnarray}
\mathcal{E}_g &=& \frac{P^2}{2} - \frac{2}{\sqrt{|x_1 (t_i)|^2 + a^2}} - \frac{2}{\sqrt{|x_2 (t_i)|^2 + a^2}}\nonumber \\
&&  + \frac{1}{\sqrt{|x_1 (t_i) - x_2(t_i)|^2 + b^2}} .
\label{eq:def_P}
\end{eqnarray}
Similar results are obtained for different slices. Figure~\ref{fig:Manifold} indicates the ionization mechanism, TS1, TS2 double ionization, or delayed or direct single ionization, undergone by the trajectory for each initial condition $(x_1(t_i),x_2(t_i))$ allowed by Eq.~(\ref{eq:def_P}). From now on, we distinguish ``the target'' from ``the atom''. Before the impact, it is clear which electron is the impact electron and which electrons belong to the target. After the impact, the impact electron might be captured by the atom. But for $\epsilon_0 \geq |\mathcal{E}_g|$, at least one electron reaches the detector. Thus, ``the atom'' refers to the ionic core and the two remaining electrons, namely those that are not the first to reach the detector. The energy of the atom $E$ is defined as the sum of the energy of the two remaining electrons. The energy conservation law imposes $\mathcal{E}_g + \epsilon_0 = E + T_{\mathrm{first}}$ where $T_{\mathrm{first}}$ is the kinetic energy of the first electron reaching the detector. Figure~\ref{fig:Energy} displays the energy of the atom $E$ at $t_f = 2000 \unites{a.u.}$, for each initial condition $(x_1(t_i) , x_2(t_i))$ of the slice of the phase space we consider. 
\par
In Fig.~\ref{fig:Manifold}, we observe clearly delimited areas associated with each single ionization mechanism with a slight predominance of the direct single ionization on this slice. The mechanism areas are intertwined in phase space in a rather complex way (more visible in the inset of Fig.~\ref{fig:Manifold}). We observe a similar intertwining in Fig.~\ref{fig:Energy}. This intertwining is mostly due to the chaotic nature of the dynamics of the target electrons before the impact. The noticeable difference between Figs.~\ref{fig:Manifold} and \ref{fig:Energy} comes from the delayed single ionization region. We observe that regions associated with the delayed ionization mechanism in Fig.~\ref{fig:Manifold} in gray, correspond to negative energy regions in Fig.~\ref{fig:Energy} in dark gray (blue online). In Fig.~\ref{fig:Energy}, these dark gray (blue online) regions are very regular in the sense that a small variation of initial conditions leads to small energy variation $E$ of the atom. However, in Fig.~\ref{fig:Manifold} we observe that the delayed single ionization region has a chaotic nature, in the sense that nearby initial conditions can lead to drastically different ionization times $\Delta t$ (see inset of Fig.~\ref{fig:Manifold}). This behavior is expected since delayed single ionization occurs by chaotic diffusion of an excited atom. In contrast, direct single ionization seems regular in the sense that nearby initial conditions lead to the same outcome in a generic way. 
\par
During delayed single ionization, the two target electrons are both bound for some time after the impact (which can be arbitrarily long). These electrons share the energy supplied by the impact electron, such that one of them describes large orbits. For instance, in Fig.~\ref{fig:delayed}, after the impact, $e_1^-$ describes large orbits while $e_2^-$ stays close to the ionic core. In this situation, in particular when $e_1^-$ is on the furthest point of the orbit, $e_2^-$ screens the charge of the ion. Since $|\br_1|$ is large, the energy of the atom is $E = E_1 + E_2$, with
\begin{eqnarray*}
E_1 &=& \frac{|\bp_1|^2}{2} - \frac{2}{\sqrt{|\br_1|^2 + a^2 }} + \frac{1}{\sqrt{|\br_1 - \br_2|^2 + b^2 }},\\
 &\simeq & \frac{|\bp_1|^2}{2} - \frac{1}{|\br_1|} ,
\end{eqnarray*}
the energy of $e_1^-$, and 
$$
E_2 = \frac{|\bp_2|^2}{2} - \frac{2}{\sqrt{|\br_2|^2 + a^2 }} ,
$$
the energy of $e_2^-$. The electron $e_1^-$ remains bounded since it comes back to the ionic core, imposing $E_1 < 0$. Moreover, $e_2^-$ also remains bounded, so $E_2 < 0$. As a consequence, delayed single ionization can occur only if the energy of the target after impact is negative, i.e. $E < 0$. This is confirmed by comparing Figs.~\ref{fig:Manifold} and \ref{fig:Energy}, where we observe that $E < 0$ for delayed single ionization. This is a necessary energy condition, but it is not sufficient, since $E<0$ can also lead to direct single ionization. 
\begin{figure}
\centering
\includegraphics[width=0.8\textwidth]{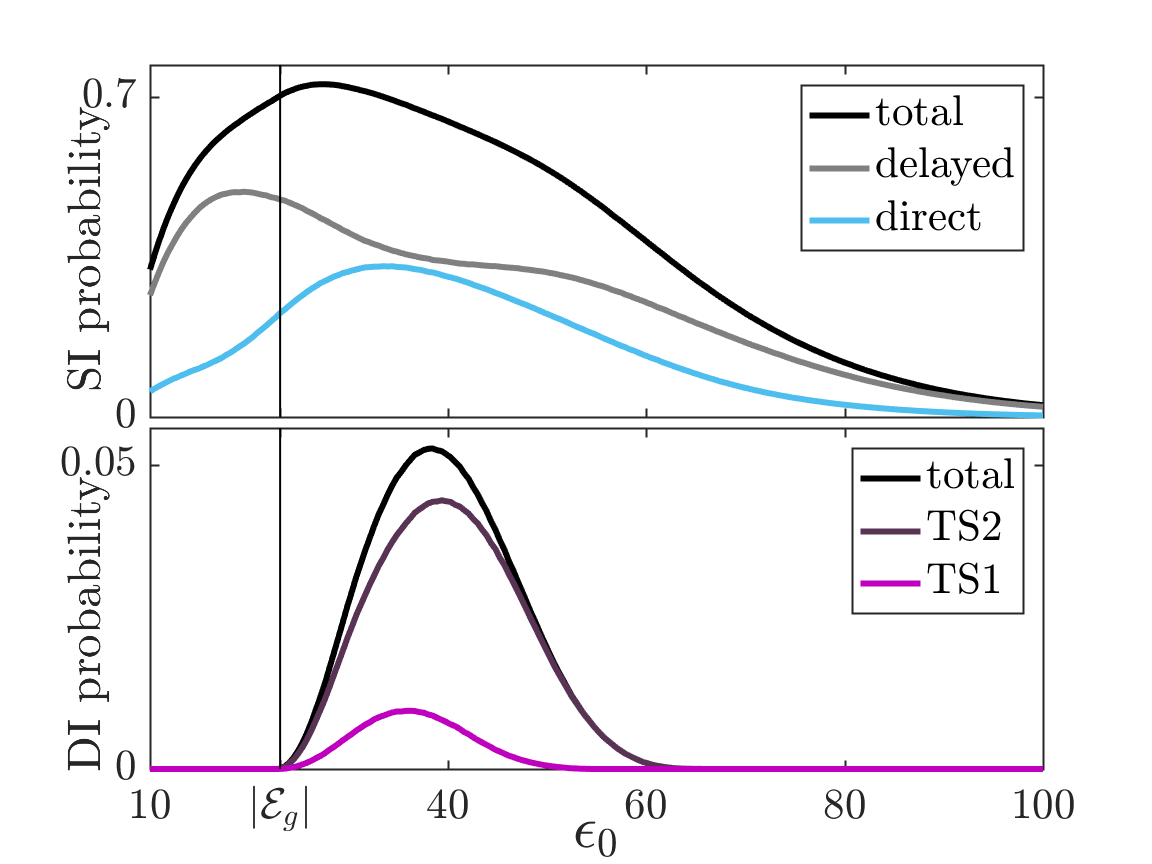}
\caption{(Color online) Upper panel: Single ionization (SI) probability as a function of the impact energy $\epsilon_0$ by delayed and direct mechanisms. Lower panel: Double ionization (DI) probability as a function of the impact energy $\epsilon_0$ by TS1 and TS2 mechanisms. The vertical line is at $\epsilon_0 = | \mathcal{E}_g |$. The integration time is $t_f = 800 \unites{a.u.}$ Here $\epsilon_0$ in $\unites{eV}$.}
\label{fig:Mechanisms}
\end{figure}
\begin{figure}
\centering
\includegraphics[width=0.8\textwidth]{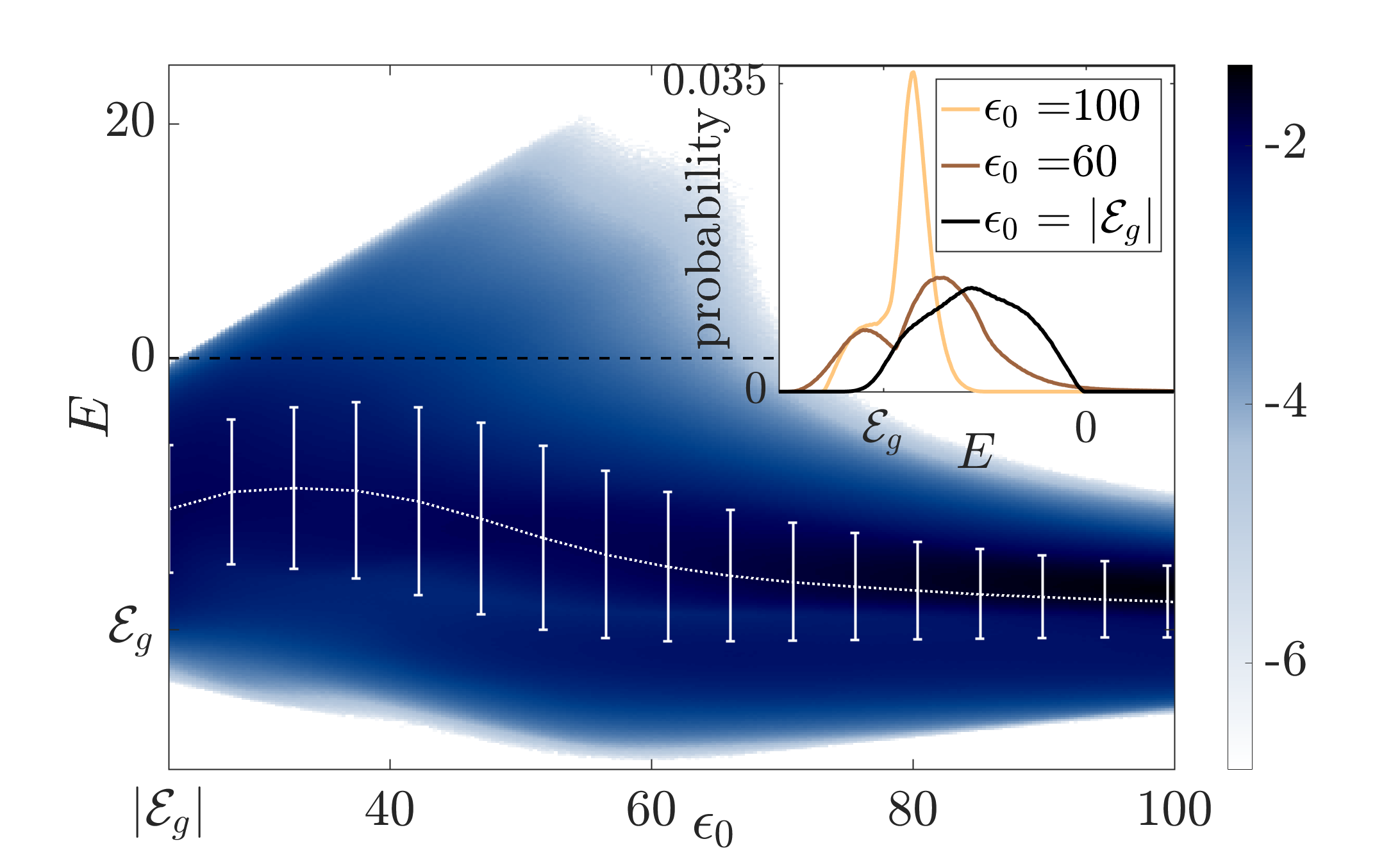}
\caption{(Color online) Probability density function of $E$ in logarithmic scale as a function of the impact energy $\epsilon_0$, where $E$ is the energy of the atom when at least one electron has reached the detector. The white line is $\langle E \rangle$, the average of $E$ for a fixed $\epsilon_0$ with the bars indicating one standard deviation. The inset shows the probability density function as a function of the final energy of the atom $E$ for fixed impact energy $\epsilon_0$. Here $E$ and $\epsilon_0$ in $\unites{eV}$.}
\label{fig:pdf}
\end{figure}
\par
Figure~\ref{fig:Mechanisms} represents the probability of each mechanism as a function of the impact energy $\epsilon_0$. Figure~\ref{fig:pdf} represents the probability density function of $E$ as a function of $\epsilon_0$, i.e. the probability that the energy of the atom long after the impact is $E$, for a given impact energy $\epsilon_0$. We have also represented $\langle E \rangle$, the average of the energy of the atom after impact for a given impact energy $\epsilon_0$. Since the mechanisms are related to energy conditions, we examine the relationships between the probability of each mechanism (Fig.~\ref{fig:Mechanisms}) and the probability density function and $\langle E \rangle$ (Fig.~\ref{fig:pdf}).
\par
In Fig.~\ref{fig:Mechanisms}, we observe that the most probable scenario for single ionization is usually delayed ionization, and its maximum is reached for $\epsilon_0 \sim 15 \unites{eV}$, similarly to the maximum of the experimental total cross section of single ionization (see Fig.~\ref{fig:SI_DI}). A dip in the delayed single ionization probability is observed when the direct single ionization channel is no longer negligible, namely in the region $\epsilon_0 \in [ | \mathcal{E}_g | , 60] \unites{eV}$. A necessary condition for a delayed single ionization process is that $E < 0$, and we observe a dominance of the delayed single ionization process for the impact energy range where the probability of having $E > 0$ is zero. However, there are no constraints for the direct single ionization process. We observe that the larger $\langle E \rangle$ is (Fig.~\ref{fig:pdf}), the larger is the direct ionization probability (Fig.~\ref{fig:Mechanisms}). This observation suggests that direct single ionization may depend on efficient energy transfer from the impact electron to the atom. In the inset of Fig.~\ref{fig:pdf}, we observe that for a constant impact energy $\epsilon_0 \in [ 50 , 80 ] \unites{eV}$, the probability density function has two distinct bumps (visible for $\epsilon_0 = 60 \unites{eV}$ in the inset of Fig.~\ref{fig:pdf}). The lowest bump in the probability density function is peaked around $E < \mathcal{E}_g$. Consequently, this bump corresponds to non-ionization. The highest bump in the probability density function is peaked around $E > \mathcal{E}_g$. This bump corresponds to single and double ionization. Moreover, this bump is wider in the region $\epsilon_0 \in [ | \mathcal{E}_g | , 40 ] \unites{eV}$, leading to larger energy transfers to the target and hence potentially more ionization.
\par
In summary, the classical model displays two mechanisms of single ionization, delayed and direct. For the delayed (indirect) single ionization, the correlation between the electrons of the atom plays a prominent role, and the mechanism involves chaotic diffusion. The necessary condition to obtain delayed single ionization is that the energy of the atom after the impact is negative, i.e. $E < 0$. Delayed ionization, also called excitation-autoionization, is the most probable process for single ionization. This is particularly true when $\epsilon_0 <  | \mathcal{E}_g |$. For direct single ionization, the energy transfer from impact electron to the target is important, while the electron-electron correlation inside the target plays a lesser role. 

\subsection{Double ionization mechanisms}
\begin{figure}
\centering
\includegraphics[width=0.8\textwidth]{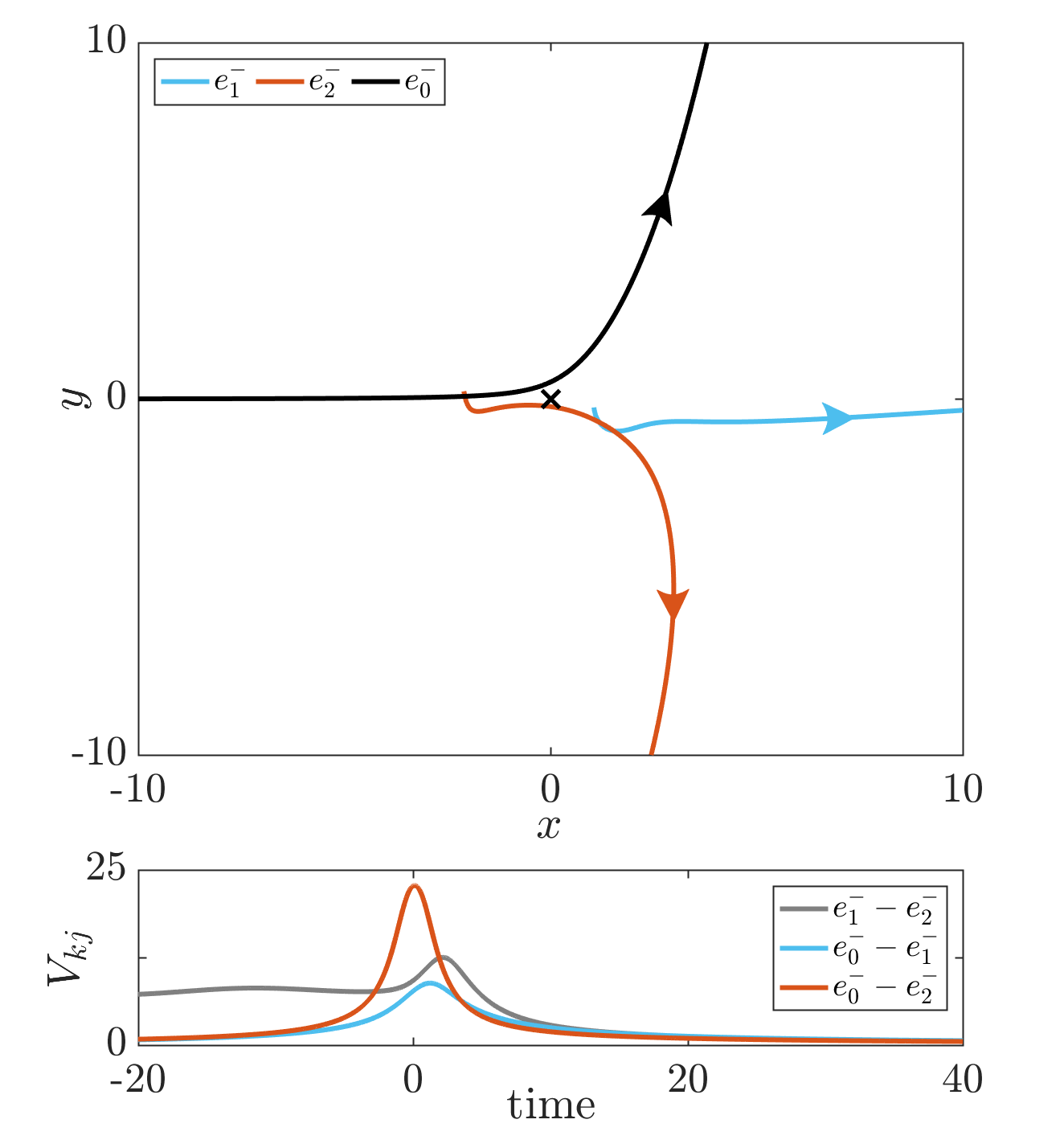}
\caption{(Color online) Upper panel: Sample trajectory undergoing TS1 in configuration space. Only the part of the trajectory after the impact is represented. Lower panel: Corresponding $V_{kj}(t)$ as a function of time for $(k,j) = (0,1)$ , $(0,2)$ and $(1,2)$. The position of the ionic core is identified by a cross. The impact energy is $\epsilon_0=40\unites{eV}$. Here $V_{kj}$ in $\unites{eV}$ and all other quantities in atomic units.}
\label{fig:TS1}
\end{figure} 
\begin{figure}
\centering
\includegraphics[width=0.8\textwidth]{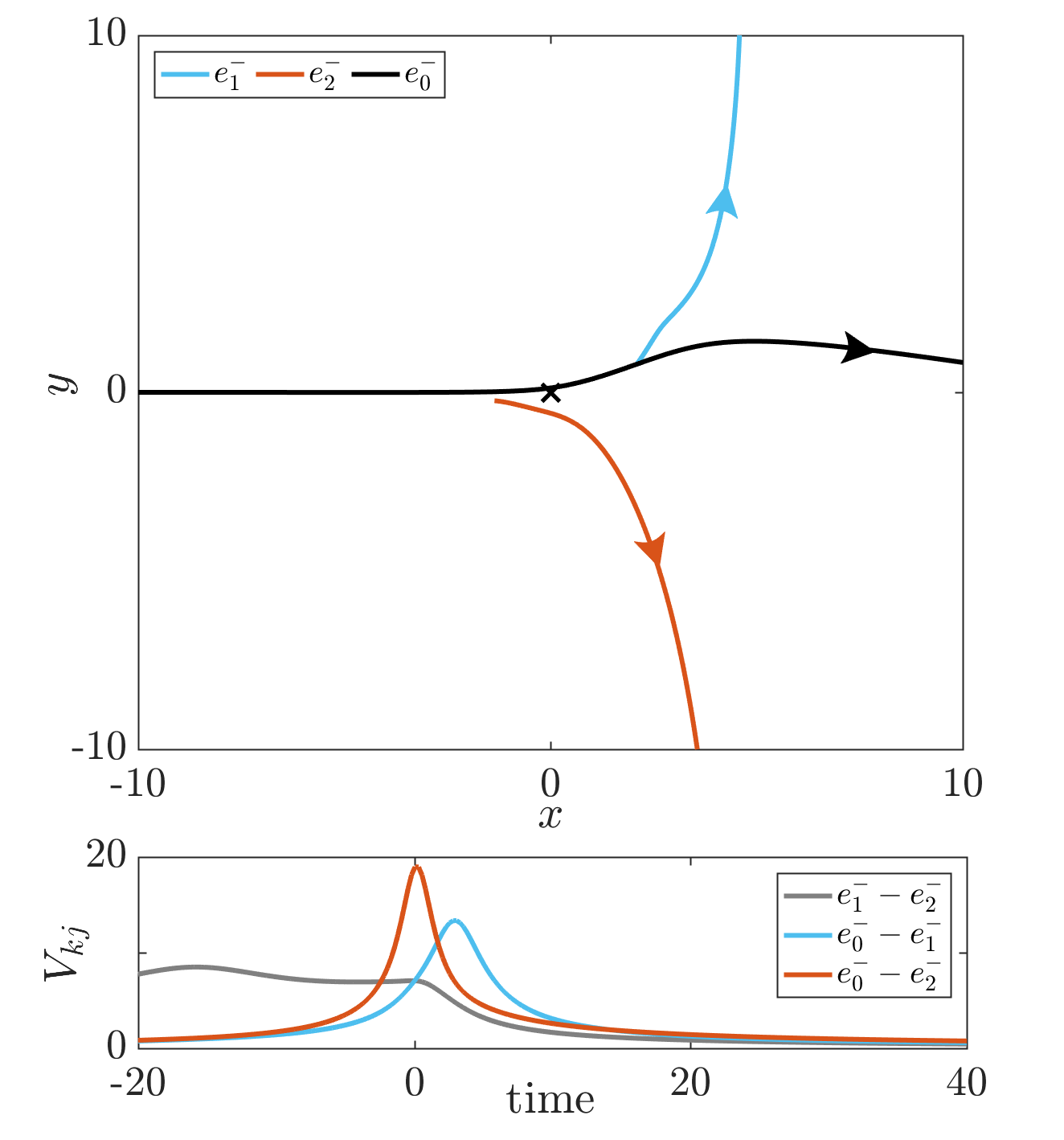}
\caption{(Color online) Upper panel: Sample trajectory undergoing TS2 in configuration space. Only the part of the trajectory after the impact is represented. Lower panel: Corresponding $V_{kj}(t)$ as a function of time for $(k,j) = (0,1)$ , $(0,2)$ and $(1,2)$. The position of the ionic core is identified by a cross. The impact energy is $\epsilon_0=40\unites{eV}$. Here $V_{kj}$ in $\unites{eV}$ and all other quantities in atomic units.}
\label{fig:TS2}
\end{figure} 
By examining  a large set of double ionizing trajectories associated with Hamiltonian~(\ref{eq:Hamiltonian}), only two distinct double ionization mechanisms have been identified in agreement with the literature: The two-step one interaction (TS1) and two-step two interaction (TS2) mechanisms \citep{Carlson1965,Gryzinski1965,Tweed1992,Gryzinski1999}.
\par
Figure~\ref{fig:TS1} represents the TS1 mechanism.  We observe peaks on the curves $V_{02}(t)$, $V_{01}(t)$, and $V_{12}(t)$ near the impact time $t\approx 0$. First, the impact electron collides with $e_2^-$ as the first peak to occur is for $V_{02}$, and subsequently the impact electron leaves the ionic core. Then, $e_2^-$ collides with $e_1^-$, and they both leave the ionic core. During the second collision, the peaks in both $V_{01}(t)$ and $V_{12}(t)$ indicate that the impact electron and $e_2^-$ are both involved in the ionization of $e_1^-$. Nonetheless, the impact electron contributes less to the ionization process during the second collision, so the dominant interaction is between the two electrons of the target. So, we consider that the impact electron has had only one interaction with the electrons of the target, and this makes it a TS1 mechanism. 
\par
In the TS2 mechanism, the impact electron interacts with both electrons of the target. Figure \ref{fig:TS2} represents the TS2 mechanism. In a similar way as for the TS1 mechanism, we observe peaks on the $V_{02}$, $V_{01}$ and $V_{12}$ interaction curves near the impact time $t\approx 0$. First, the impact electron collides with $e_2^-$, and $e_2^-$ leaves the ionic core. Subsequently, the impact electron collides with $e_1^-$, and both leave the ionic core. 
\par
Comparing the local maxima of $V_{01}$ and $V_{02}$ around $t=0$ (time of impact), we numerically discriminate trajectories belonging to TS1 or TS2. In Fig.~\ref{fig:Manifold}, we observe that double ionizations occur in highly localized regions in phase space (at least in the slice we consider in Fig.~\ref{fig:Manifold}). Moreover, the TS1 and TS2 regions are intertwined, i.e. it is difficult to predict which mechanism will be involved for a given initial condition. Also on the slice depicted on this figure, we have found that the number of TS2 trajectories is roughly three times larger than that of TS1. 
\par
The atom is doubly ionized if the positions of the electrons are such that $| \br_k | \to \infty$ for all $k=0,1,2$ as $t \to \infty$, i.e. the interaction between the ionic core and the electrons is negligible. Consequently, Hamiltonian~(\ref{eq:Hamiltonian}) reduces to
$$
H =   \sum_{k=0}^2 \frac{|\bp_k|^2}{2} + \sum_{j>k} \frac{1}{\sqrt{|\br_k - \br_j|^2 + b^2}} \geq 0,
$$
irrespective of the relative positions of the three electrons.
Using Eq.~(\ref{eq:total_energy}), a necessary energy condition to obtain double ionization is
$$
\epsilon_0 + \mathcal{E}_g \geq 0,
$$
as confirmed by the double ionization probability in Fig.~\ref{fig:SI_DI}. 
During an ionization process, electrons have a tendency to escape with large relative distances because of electron-electron repulsion, i.e. $|\br_k - \br_j| \to \infty$. Consequently, the interaction between the electrons vanishes and the Hamiltonian can be decomposed as the sum of the final three kinetic energies, denoted $T_k = |\bp_k|^2/2$, such that
\begin{equation}
\epsilon_0 + \mathcal{E}_g = T_0 + T_1 + T_2 .
\label{eq:before_after}
\end{equation}
Since $T_k \geq 0$, the final energy of each electron does not exceed the total energy of the system, i.e.
\begin{equation}
T_k \leq \epsilon_0 + \mathcal{E}_g ,
\label{eq:condition2}
\end{equation}
for all $k=0,1,2$. In particular, because of inequality~(\ref{eq:condition2}) and since $E$ is composed of the sum of the energy of two electrons, Eq.~(\ref{eq:before_after}) leads to $E \geq 0$. This is confirmed by comparing Figs.~\ref{fig:Energy} and \ref{fig:Manifold}, where we observe that the double ionization regions on Fig.~\ref{fig:Manifold} correspond to dark gray (dark red online) regions on Fig.~\ref{fig:Energy}, representing $E \geq 0$. Double ionization occurs only if the final energy of the target atom is positive, which happens, as we observe on Fig.~\ref{fig:pdf}, only in the region $\epsilon_0 \in [ | \mathcal{E}_g | , 70] \unites{eV}$. We notice that this range contains the range where double ionization probability is non-zero (see Figs.~\ref{fig:SI_DI} and \ref{fig:Mechanisms}). From Fig.~\ref{fig:pdf}, we see that the maximum of the average energy of the atom $\langle E \rangle$ after impact is obtained for $\epsilon_0 \sim 40 \unites{eV}$, which is consistent with the impact energy leading to the maximum double ionization probability (see Fig.~\ref{fig:SI_DI}). 
%For $\epsilon_0 > 45\unites{eV}$, the probability density function is shrunk around $-\mathcal{E}_g$ for increasing impact energy. Indeed, larger is the impact energy, and shorter is the interaction time (i.e., the time the impact electron spends close to the target atom). 
\par
In the lower panel of Fig.~\ref{fig:Mechanisms}, we observe that for any impact energy, double ionization is clearly dominated by the TS2 mechanism. The TS1 mechanism occurs in the range $\epsilon_0 \in [ | \mathcal{E}_g |  , 45] \unites{eV}$, while TS2 occurs in a broader range of impact energies $\epsilon_0 \in [ | \mathcal{E}_g | , 60] \unites{eV}$, which is the range of impact energy where double ionizations are detected. So here again, the results of Figs.~\ref{fig:SI_DI} and \ref{fig:pdf} are consistent with the energy conditions.
\par
In summary, the classical model~(\ref{eq:Hamiltonian}) displays only two double ionization mechanisms which are direct, ruling out indirect mechanisms like excitation-autoionization for our two-active electron model. For the TS1 mechanism, the correlation between the electrons of the target atom is largely involved, whereas this correlation is neglected in the TS2 mechanism. A necessary condition to obtain double ionization is that the energy of the atom after the impact is positive, i.e. $E \geq 0$. The TS2 mechanism is the dominant scenario for double ionization. This is in agreement with the experimental results of Refs.~\citep{LahmamBennani2010,Casagrande2011,Li2011,Li2012}. 

\subsection{Robustness of the results \label{sec:robustness}}
\begin{figure}
\centering
\includegraphics[width=0.8\textwidth]{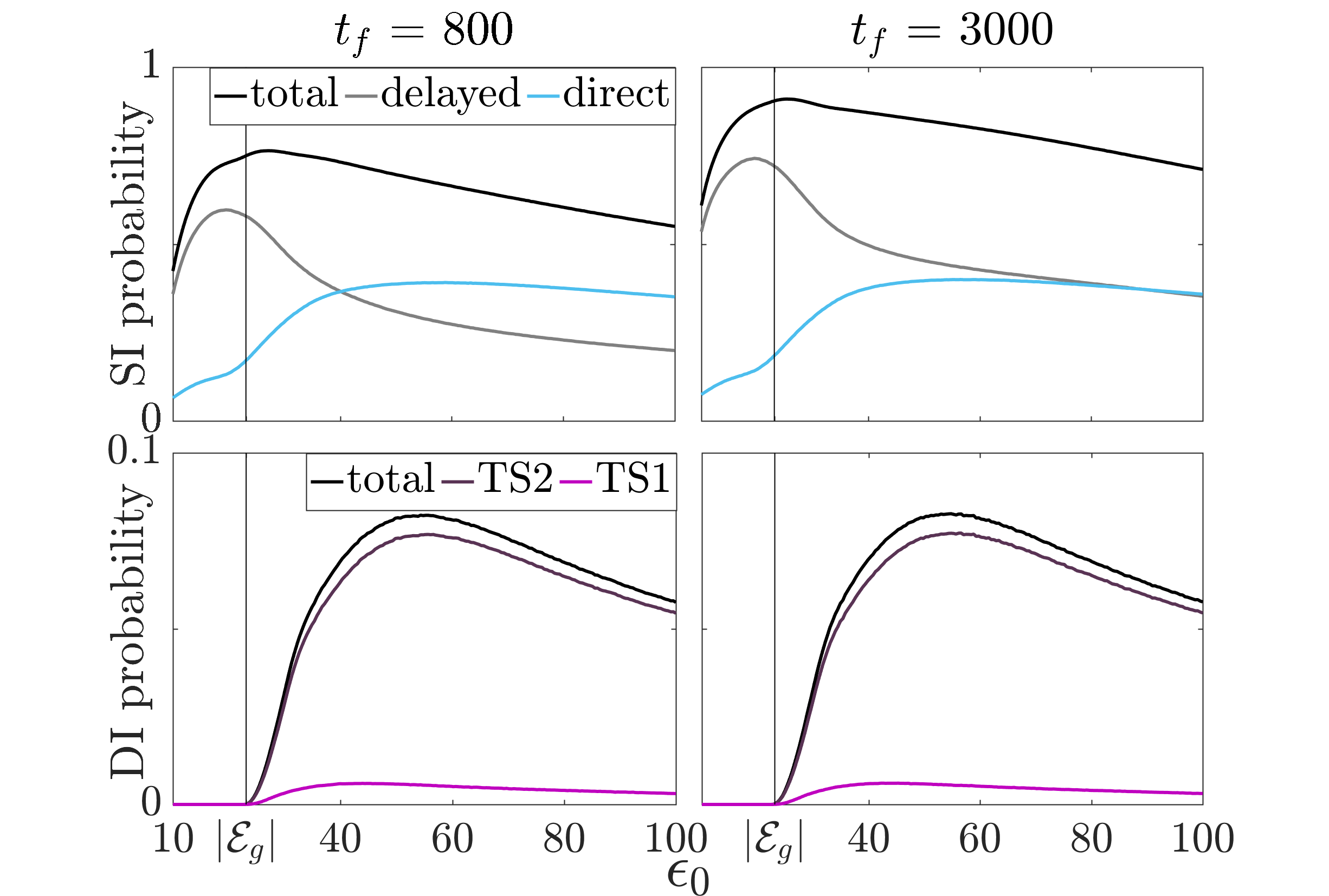}
\caption{(Color online) Upper panels: Total single ionization (SI) probability as a function of the impact energy $\epsilon_0$ and its decomposition in delayed and direct mechanisms. Lower panels: Total double ionization (DI) probability as a function of the impact energy $\epsilon_0$ and its decomposition in TS1 and TS2 mechanisms. Single and double ionization both computed with $b = 0.3$. Left panels: The measurement is performed at $t_f = 800 \unites{a.u.}$ after the impact. Right panels: The measurement is performed at $t_f = 3000 \unites{a.u.}$ after the impact. The vertical line is at $\epsilon_0 = | \mathcal{E}_g |$. Here $\epsilon_0$ in $\unites{eV}$ and all other quantities in atomic units.}
\label{fig:mechanims_b}
\end{figure} 
\begin{figure}
\centering
\includegraphics[width=0.8\textwidth]{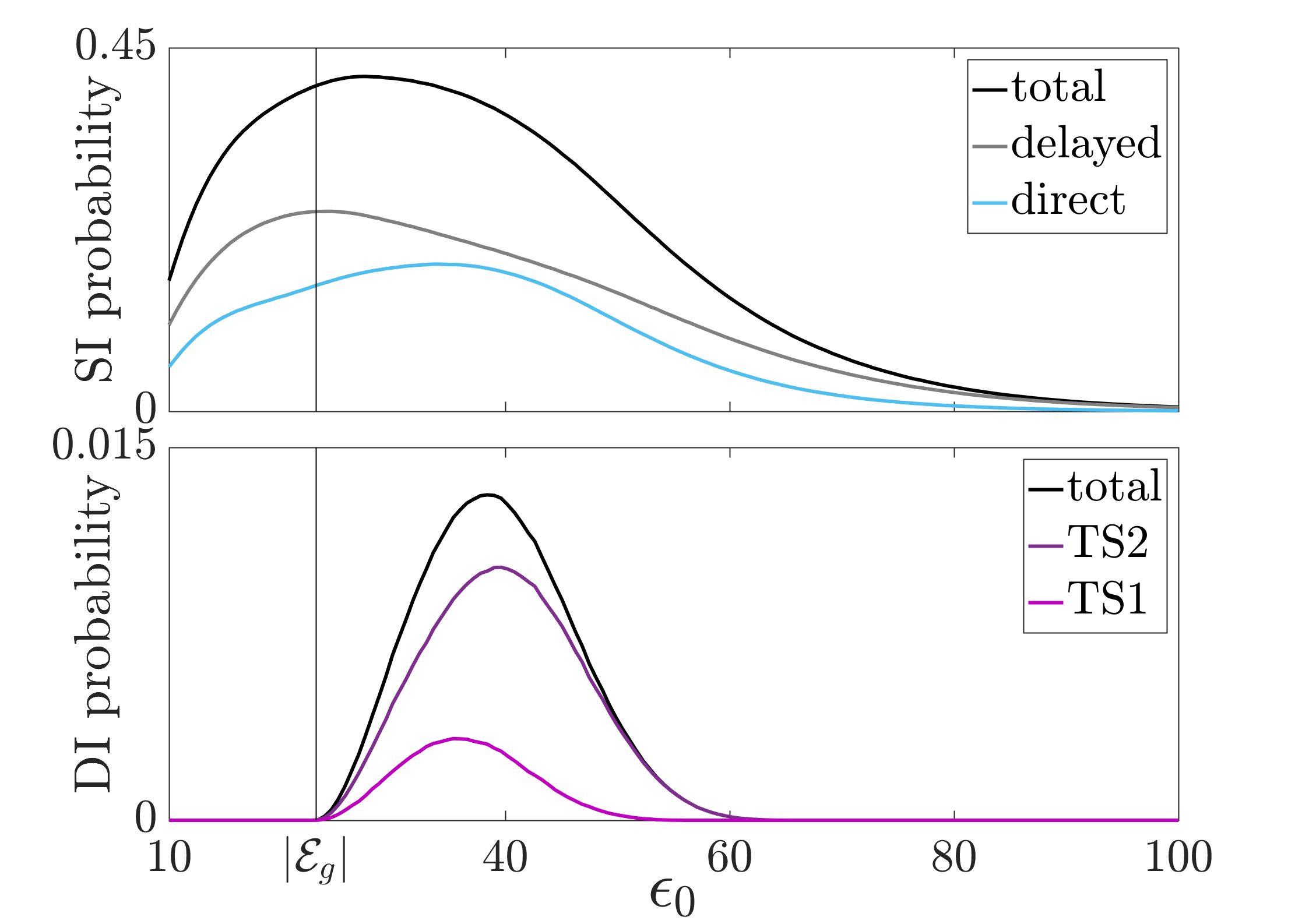}
\caption{(Color online) Upper panels: Total single ionization (SI) probability as a function of the impact energy $\epsilon_0$ and its decomposition in delayed and direct mechanisms. Lower panels: Total double ionization (DI) probability as a function of the impact energy $\epsilon_0$ and its decomposition in TS1 and TS2 mechanisms. Single and double ionization both computed with $b = 1 \unites{a.u.}$ and $t_f = 800 \unites{a.u.}$ For each trajectory, the impact parameter $y_0(t_i)$ is chosen randomly in a range $[0 , 3] \unites{a.u.}$ Here $\epsilon_0$ in $\unites{eV}$.}
\label{fig:mechanisms_ip}
\end{figure} 
The question we address now is how generic and robust these results are, in particular, with respect to some parameters of the model~(\ref{eq:Hamiltonian}): Softening parameter $b$, impact parameter $y_0(t_i)$, and integration time $t_f$. Figure~\ref{fig:mechanims_b} shows the probability of each ionization mechanism for $b = 0.3\unites{a.u.}$, i.e. for a stronger electron-electron interaction than on Fig.~\ref{fig:Mechanisms}, measured for $t_f = 800 \unites{a.u.}$ and $t_f = 3,000 \unites{a.u.}$ Figure \ref{fig:mechanisms_ip} shows the probability of each ionization mechanism, where for each trajectory the impact parameter $y_0(t_i)$ is chosen randomly in the range $[0 , 3]\unites{a.u.}$
\par
Comparing Figs.~\ref{fig:Mechanisms} and \ref{fig:mechanims_b}, we observe that decreasing the softening parameter $b$ increases the double ionization probability as expected, and extends the range of impact energy $\epsilon_0$ where double ionization is observed. The behavior before the maximum double ionization probability is similar, but after the maximum, the decrease is smoother, as observed in the experimentally measured total double ionization cross sections. The TS2 mechanism is even more dominant than for $b=1\unites{a.u.}$ in comparison with the TS1 contribution. Concerning the single ionization mechanisms, we notice that the contribution of the direct single ionization is increased and even dominant for large values of the impact energy, in contrast with the $b = 1\unites{a.u.}$ case. However, we will see below that this is an artefact of a final integration time $t_f$. Moreover, we have seen that for decreasing $b$, the intertwining observed in Figs.~\ref{fig:Manifold} and \ref{fig:Energy} is more regular (not pictured in this article). In contrast, if $b$ increases, we observe an increase of the sensitivity to initial conditions, and the appearance chaotic sea.  
\par
Comparing Figs.~\ref{fig:Mechanisms} and \ref{fig:mechanisms_ip}, we observe that if $y_0 (t_i) \neq 0$, the ionization probabilities decreases significantly, but the qualitative feature of the curves remains almost identical. The influence of the impact parameter on the probability curves is only quantitative. Taking $y_0 (t_i) = 0$ allows a higher probability to obtain ionization.
\par
Finally, we observe that the integration time $t_f$ does not influence the double ionization probability and the direct single ionization curves. Indeed, because delayed ionization processes can take an arbitrary long time, the longer the integration time $t_f$ is, the higher the delayed ionization probability will be. We notice that delayed ionization becomes more dominant. Furthermore, increasing $t_f$ moves the maximum of the single ionization probability to the left. However, for sufficiently long integration times (e.g. $1500 \unites{a.u.}$) the delayed single ionization converges, so that considering larger integration times is unnecessary for practical purposes. 
\par
In summary, the softening parameter $b$, the impact parameter $y_0(t_i)$, and the integration time $t_f$ influence some features of the ionization probability curves, like for instance the amount of single and double ionization. However, some features remain unchanged, like the predominance of delayed single ionization and of TS2 double ionization. In terms of the mechanisms, the analysis we presented is robust with respect to changes in these parameters.

\section*{Conclusions}
In summary, we proposed a two-active electron classical Hamiltonian model for the $(e,2 e)$ and $(e,3e)$ processes for a target atom with two loosely bound electrons, which reproduces notable features of the experimentally measured single and double ionization cross sections. We have shown that a two-dimensional model is capable of capturing some features of the experimentally measured curves. The classical approach allowed us to identify the mechanisms involved in the ionization processes by examining trajectories in phase space, and their relative contributions by analyzing families of trajectories. Four mechanisms have been observed: Direct and delayed single ionization, and the TS1 and TS2 mechanisms. The behavior of these mechanisms as a function of the impact energy $\epsilon_0$ has been studied, as well as where they occur in phase space. The delayed single ionization displays a strong dependence with respect to initial conditions, and is associated with single ionization by chaotic diffusion. The TS2 mechanism is found to dominate in the double ionization processes, in agreement with existing experiments. 
\par
The significant disagreement between the double ionization probability from Hamiltonian~(\ref{eq:Hamiltonian}) and the experimentally measured double ionization cross section \citep{McCallion1992a} suggests a large contribution of the inner shell electrons in the double ionization processes for $\epsilon_0 > 55 \unites{eV}$. A two-active electron model can not reproduce this feature, nor reproduce fully the knee observed in the double ionization cross section. However, a two-active electron model is able to capture the behavior in the low energy region, the first part of the knee.

\section*{Acknowledgments}

The research leading to these results has received funding from the People Program (Marie Curie Actions) of the European Union’s Seventh Framework Program No. FP7/2007-2013/ under REA Grant No. 294974.
J.D. thanks the A*MIDEX financial support (No. ANR-11-IDEX-0001-02) of the French Government ``Investissements d\textquoteright Avenir'' program.
S.A.B.\ and T.U.\ acknowledge funding from the NSF (Grant No. PHY1304741).
This material is based upon research supported by the Chateaubriand Fellowship of the Office for Science \& Technology of the Embassy of France in the United States.
We acknowledge useful discussions with Maxime Perin and Fran\c{c}ois Mauger.

%\bibliographystyle{apsrev4-1}
%\bibliography{biblio_v1}


\end{document}